\documentclass[aps,prl,showpacs,twocolumn,groupedaddress]{revtex4}
\usepackage{subfigure}
\usepackage{amsmath}
\usepackage{graphicx}
\usepackage{rotating}
\usepackage{amsfonts}
\usepackage{amssymb}%
\providecommand{\U}[1]{\protect\rule{.1in}{.1in}}
\begin{document}

\title{Microscopic origin of magnetism and magnetic interactions in ferropnictides}
\author{M.D. Johannes, I.I. Mazin}
\affiliation{Code 6393, Naval Research Laboratory, Washington, D.C. 20375}
\date{Printed on \today}

\begin{abstract} One year after their initial discovery, two schools of thought have crystallized 
regarding the electronic structure and magnetic properties of ferropnictide systems. One postulates 
that these are itinerant weakly correlated metallic systems that become magnetic by virtue of 
spin-Peierls type transition due to near-nesting between the hole and the electron Fermi surface 
pockets. The other argues these materials are strongly or at least moderately correlated, the electrons 
are considerably localized and close to a Mott-Hubbard transition, with the local magnetic moments 
interacting via short-range superexchange. In this paper we argue that neither picture is fully 
correct. The systems are moderately correlated, but with correlations driven by Hund's rule coupling 
rather than by the on-site Hubbard repulsion. The iron moments are largely local, driven by Hund's 
intra-atomic exchange. Superexchange is not operative and the interactions between the Fe moments are 
considerably long-range and driven mostly by one-electron energies of all occupied states.

\end{abstract}

\pacs{Pacs Numbers: }
\maketitle

Ferropnictides are still attracting widespread attention from researchers both inside and outside the 
field of superconductivity. There is now a nearly universal agreement that magnetism and, specifically, 
proximity to an antiferromagnetic \textquotedblleft stripe-order\textquotedblright\ transition plays a 
major role in the physics of these compounds. There is also growing evidence that the magnetic properties 
and correlation effects in this system are not controlled by the Hubbard $U$ as in cuprates (spectroscopy 
tells us that Hubbard correlations are weak and the effective $U$ is on the order of 1 eV\cite{PES}, 
smaller than the bandwidth; first principles calculations of $U$ support this\cite{U}). On the other hand, 
the multiband character of Fe bands and the large intra-atomic (Hund's) exchange coupling in Fe suggest 
that the 
Hund's $J$ may play the main role in the magnetism..

As opposed to the Hubbard $U,$ the Hund's $J$ is generally well accounted for in density-functional 
calculations (where it is called the Stoner $I$).) Indeed, the local density approximation (LDA) 
correctly predicts the particular antiferromagnetic and structural ground state of undoped 
ferropnictides, in striking contrast to the cuprates. In view of this, it is instrumental to trace down 
the origin of magnetism $within$ LDA, and to disentangle the nature of magnetic interactions captured by 
this approach. It is highly likely that the physics uncovered by density functional theory will reflect 
the actual physics of these systems. Given the heated (but largely devoid of solid facts) discussion of 
whether antiferromagnetism in pnictides is due to Fermi surface nesting or to second neighbor 
superexchange (See Ref. \cite{MS} for a review), a clear understanding of, at least, the message that 
LDA calculations send seems highly necessary.

In this paper, we analyze the magnetic interactions and demonstrate that
neither of the above two views (often presented as an axiomatic dilemma) is
correct. The magnetism appears due to $local$ Hund's rule coupling, while the
particular ground state is selected by itinerant, essentially one-electron
interactions, of which only a small part is
played by the Fermi surface nesting. We will also explain why conventional
\textquotedblleft Anderson-Kanamori\textquotedblright\ superexchange is not
operative here, and will show some striking examples where calculations
and experiment contradict both local superexchange and spin-Peierls pictures,
yet are perfectly understandable on the basis of one-electron energy balance.

We start with a qualitative analysis. The Hund's rule coupling energy in density 
functional theory is expressed as $E_{H}=-Im^{2}/4,$ where $m$ is the magnetization 
of an Fe ion, and $I\approx0.8$ eV is the Stoner factor for Fe (0.9 eV in GGA). 
Depending on the material, the self-consistent magnetic moment on Fe appears to be 
between 1.5-2 $\mu_{B}$ in LDA and 1.8-2.5 in GGA. The corresponding energy gain even 
in LDA is 0.5 eV, which is remarkably large. In other words, every individual Fe 
wants to be strongly magnetic and the advantage of spin polarization should lead to a 
magnetic ground state at the mean field level, unless an unusually large kinetic 
energy penalty exists. However, this is exactly the case for the formation of a 
ferromagnetic configuration. To create a magnetization $m$ on Fe, one needs to move 
approximately 1.15 (to account for the relative share of Fe-d orbitals at the Fermi 
level) spin-minority electrons into unoccupied spin-majority states, incurring an 
energy loss of $\approx(1.15m)^{2}/N_{\uparrow}(E_{F}).$ The density of states (DOS) 
per Fe, $N_{\uparrow}(E_{F}),$ varies between 1 and 1.5 eV$^{-1},$ depending on the 
system, creating an energy loss for $m=1.5$ $\mu_{B}$ of 0.5-0.8 eV. This cost is 
about as large as the Hund's rule energy gain estimated above.  This shows that the 
system is on the verge of a ferromagnetic instability, but nothing more.

In low-DOS metals, magnetization without a large cost in kinetic energy is possible if some type of 
antiferromagnetic arrangement is formed (cf. metal Cr and Mn). For a broad band metal, this narrows the 
conductivity band, but as long as the exchange splitting is smaller than the bandwidth, the cost is small. 
Because in ferropnictides the calculated bandwidth is 5-6 eV and the exchange splitting $mI$ is at most 2 
eV, this mechanism should be very favorable.

It is interesting to consider how the system determines which particular
AFM arrangement is most profitable from the point of view of the one-electron energy
(note that LDA calculations can be forced to converge to nearly any AFM
pattern, but not to a ferromagnetic state). If the resulting magnetization is
small, the answer is obvious: the second derivative of the total energy with
respect to magnetization is defined by the noninteracting susceptibility at
the AF wave vector \textbf{Q}, $\partial^{2}E/\partial m^{2}=-\chi
_{0}^{-1}(\mathbf{Q)}$ (with the small caveat that an actual spin density wave
is not a single harmonic, but includes all wave vectors \textbf{Q+G, }where
\textbf{G} is a reciprocal lattice vector). The imaginary part of $\chi_0$ is
directly related to Fermi surface nesting, being defined, in the
constant matrix elements approximation, as $\sum_{ij}\int\delta(\varepsilon
_{\mathbf{k}i})\delta(\varepsilon_{\mathbf{k+Q,}i})d\mathbf{k,}$ while the
(actually relevant) real part collects information from all states and may or
may not have any relation to the nesting conditions (for a detailed discussion
see Ref. \cite{CDW}). 

Geometrical nesting, as a property of the Fermi surface, becomes even more disconnected from a real 
instability in the strongly nonlinear regime, $m\gtrsim1$ $\mu_{B}$, which is the case for ferropnictides. 
Monitoring the evolution of the electronic bands with increasing spin polarization\cite{FSM}, one observes 
that at $m\sim1$ $\mu_{B}$ the resulting bands can in no way be described as anticrossing downfolded 
nonmagnetic bands with partial gapping of the Fermi surface.  Rather, the entire Fe d band is fully 
restructured. Although the lowest-energy AFM state wave vector indeed coincides with the quasi-nesting 
wave vector in some cases, it is not always true, as exemplified by the case of FeTe that we discuss 
later.

It should be noted that while quasi-nesting is not particularly relevant for
the long-range ordering in the undoped crystals, it does define the low-energy excitations in
non-magnetic phases and these can perfectly well mediate superconductivity.



\begin{figure} \includegraphics[width = 0.95\linewidth]{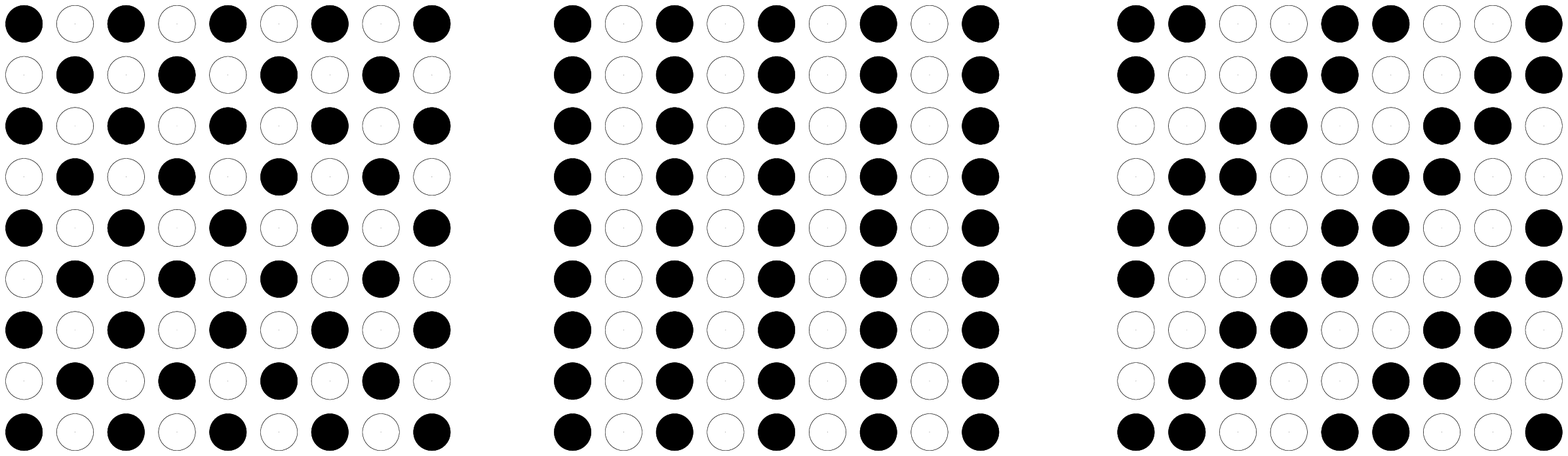} \caption{(color online) Top-down view of the 
a) checkerboard b) stripe and c) doublestripe magnetic patterns for a single FeAs or FeTe layer.  The light 
colored sites have majority up spin and the darker sites have majority down spin.} 
\label{structures} 
\end{figure}

Having established a general 
framework, we now address specific examples. First, we investigate checkerboard, stripe, and double-stripe 
magnetic structures (See Fig. \ref{structures}) and show that the stripe order is lower in energy than either 
the checkerboard or the double-stripe structure for the 122 systems, but not for FeTe. We 
use BaFe$_{2}$As$_{2}$ as an example, but the results for LaFeAsO are very similar. Our calculations were 
performed using an all-electron, full-potential LAPW package WIEN2k, in the Generalized Gradient Approximation, 
similar to Ref. \cite{PRB}.  All structures were fully relaxed (except where stated otherwise) 
using the Vienna Ab-Initio Simulation Program (VASP) \cite{vasp}, with the PAW formulation \cite{PAW} and 
also using GGA. In Table 1 we show the magnetic stabilization energies of the three different antiferromagnetic 
structures. 

\begin{table}[tbp]
\caption{Stabilization energies for various magnetic configurations in the 122 and FeTe systems.  All energies are per Fe atom.}
\begin{tabular}{|l||c|c|c|}
\hline
& checkerboard & stripe & double stripe \\
\hline
BaFe$_2$As$_2$ & 16 meV & 94 meV & 0.6 meV     \\
FeTe & --& 207 meV  & 230 meV \\
\hline
\end{tabular}
\label{stable}
\end{table}

\begin{figure} \includegraphics[height= 0.95\linewidth, angle=270]{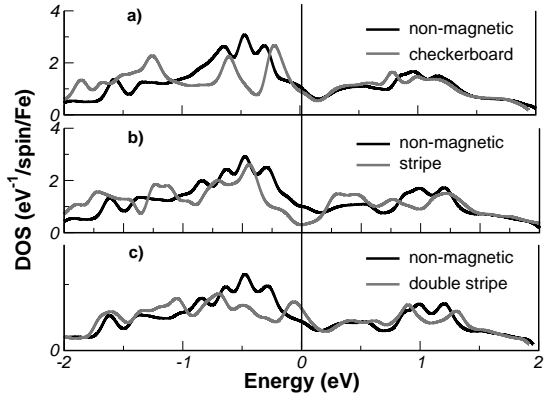} \caption{(color online) The densities 
of states for BaFe$_2$As$_2$ in the non-magnetic configuration in comparison to a) checkerboard magnetic pattern 
b) stripe (ground state) magnetic pattern and c) double stripe magnetic pattern} \label{DOS122} \end{figure}

In Fig. \ref{DOS122}a,b,c, we show the DOSs for BaFe$_2$As$_2$ in each of the three magnetic configurations 
along with the nonmagnetic DOS. Compared to the nonmagnetic DOS, we see that the checkerboard pattern has a very 
similar spectrum at and near the Fermi energy and gains one-electron energy by shifting spectral weight from the 
region between -0.5 and -1.0 downward to the region between -1.0 and -2.0.  The doublestripe pattern actually 
incurs an energy penalty at and just below E$_F$, but gains energy by shifting weight downward from between -0.2 
and -0.7 to between -1.0 and -2.0.  The ground state configuration, in contrast to the other two, gains energy 
all the way from E$_F$ to -0.9 by shifting weight downward.  This is accomplished through the opening of a large 
pseudogap (this terminology has no connection with the pseudogap in cuprates and simply signifies a depression 
in one-electron DOS around the Fermi level).  Though all three magnetic configurations are stable with respect to a 
nonmagnetic state, it is visibly the case that the stripe ordering has the greatest one electron energy 
advantage.  This is reflected in the much larger gain in total energy (See Table \ref{stable}).

Let us now compare the results with the same calculations for FeTe. As indicated in a number of papers, 
FeTe is always formed with an excess Fe, so the fact that experiment gives the double stripe structure as 
the low-temperature ground state \cite{Li} should be taken $cum$ $grano$ $salis.$ However, as Table 1 
shows, it is definitely the stoichiometric ground state in density functional calculations, and this is 
the only thing that matters for our analysis \cite{note1}.  We note here that we do not fully relax the 
FeTe structure, but only relax the internal positions.  As before these relaxations are done separately 
for magnetic and nonmagnetic cases.

\begin{figure}
\includegraphics[height= 0.95\linewidth, angle=270]{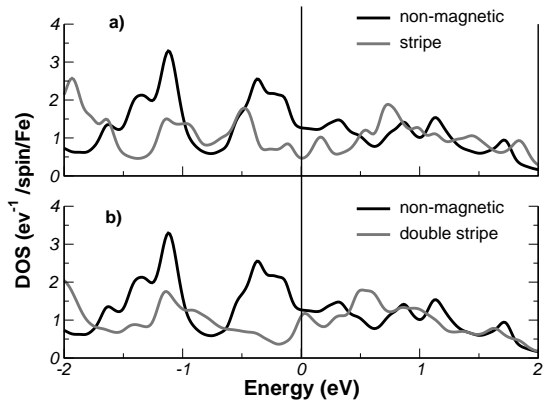}
\caption{(color online) The densities of states for FeTe in the non-magnetic 
configuration 
in comparison to a) stripe magnetic pattern and b) double stripe magnetic pattern.}
\label{FETEDOS}
\end{figure}

Table \ref{stable} indicates that for FeTe, as opposed to BaFe$_{2}$As$_{2},$ the energies of the 
single and double stripe phases are relatively close. This suggests that the crude method of 
determining the ground state by looking at the DOS may not work here, as the DOSs for the two AFM phases 
will probably be similar. Indeed, this is what we see in Fig. \ref{FETEDOS}a,b where a large downward 
shift of spectral weight is visible for both patterns. Interestingly, the nonmagnetic Fermi surfaces in 
FeTe are extremely similar to those in the 122 and 1111 materials, whereas nonmagnetic DOS and the 
magnetic electronic structure are quite different. This reinforces that FS nesting, which would be 
nearly identical for BaFe$_2$As$_2$ and FeTe, is not driving the magnetic order.  

Our calculations also provide a strong argument against superexchange.  Looking at the patterns in Fig. \ref{structures}, it is easy to see that for both the 
stripe and double stripe patterns, the first neighbor exchange, $J_1$, does not contribute, due to equal numbers of aligned and anti-aligned spins.  The second 
neighbor exchange, $J_2$, would have to be stronger than $J_1/2$ in order for the stripe pattern to be 
energetically favorable over the checkerboard pattern in a 
superexchange picture.  It has often been argued that this situation is not unreasonable since the Fe-As-Fe paths available for $J_1$ and $J_2$ are similar. For 
double stripe order in the FeTe system, however, both $J_1$ and $J_2$ cancel, leaving only $J_3$ to establish the ordering.  Considering the remarkably strong 
stability (compared to non-magnetic) calculated for double stripe order (See Table \ref{stable}), this is hard to rationalize.  Furthermore, the energy term for 
stripe is $J_2$ - $J_3$ (compared to $J_3$ alone for double stripe).  For double stripe to stabilize, 
$J_3$ could be no smaller than $J_2/2$, but the "similar 
hopping paths" justification used for $J_1$ and $J_2$ and is not available: the third neighbor exchange path is more than twice as long as the second neighbor 
one and involves As-As hopping across a distance of a full lattice constant.  Thus, the existence and stability of the double stripe order severely strains the 
credibility of the superexchange picture.  This is, in fact, to be expected since superexchange is not 
efficient when the bandwidth is much larger than the energy cost of flipping an electron's spin, which is 
precisely the case here.

This does not, however,  mean that one cannot map the dependence of the total energy onto a
suitable short-range exchange model. In fact, it is hard to imagine a case in which
this would not be possible. Yet, in carrying out this procedure for ferropnictide systems,
one should be aware of the following caveats:

(1) There is no microscopic justification (as for instance in the Hubbard
model) for introducing any $J-t$ (or $J_{1}-J_{2}-t)$ Hamiltonian.

(2) There is no guarantee that this kind of mapping can be stopped at first or second neighbors. In fact, 
accurate calculations show that at least some of the exchange parameters in these mappings decay as 
$1/R^{3},$ just as in metal iron\cite{Y-A}.

(3) The resulting exchange parameters strongly and qualitatively depend
on the long-range order established in the system. In particular, the parameters
that can be used to describe the ordered state cannot be used to describe the spin
fluctuations, and vice versa. (See Ref. \cite{Y-A} and references therein.)

(4) In the absence of superexchange, there is no reason to believe that the
total energy can be mapped onto a Heisenberg model, $\sum_{ij}J_{ij}
\mathbf{S}_{i}\cdot\mathbf{S}_{j}.$ In fact, direct calculations show that at
least one biquadratic term needs to be added to map the total energy onto the
mean-field Hamiltonian, $\sum K(\mathbf{S}_{i}\cdot\mathbf{S}
_{i+\mathbf{1}})^{2},$ where $K\sim J$. \cite{Ole}

\begin{figure}
\includegraphics[width=0.95 \linewidth]{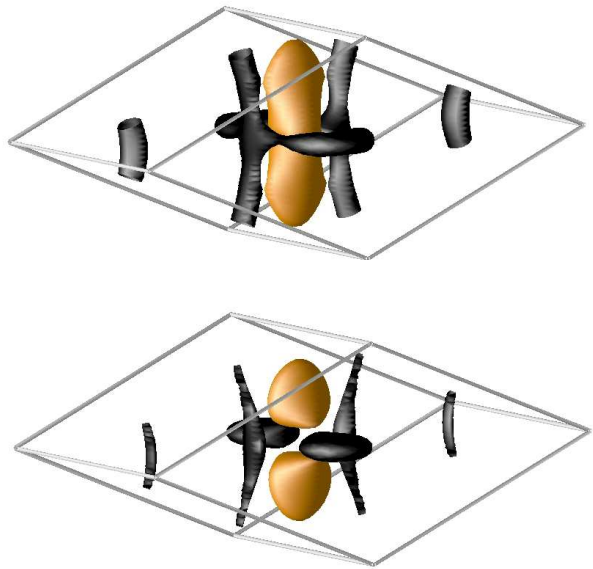}
\caption{The Fermi surfaces of stripe-ordered BaFe$_2$As$_2$.  Top panel shows the 'reverse distortion' in which the Fe-Fe distance is lengthened along the like spin direction and shortened along the unlike spin direction.  
The bottom panels shows the fully relaxed calculation which reproduces the experimentally observed distortion (to within a few percent).}
\label{FS122}
\end{figure}

We now switch our attention to the structural transition observed simultaneoulsy with the magnetic one in 
the 122 systems. Density functional calculations very accurately reproduce the experimentally observed 
distortion in which Fe ions along the stripe direction are closer to one another than Fe atoms belong to 
adjacent stripes \cite{yildirim, jesche}. We investigated whether the structural distortion, like the 
magnetic ordering, can be understood in terms of one electron energies by calculating the DOS for a 
variety of small changes in the $a$ and $b$ lattice constants.  In contrast to changing the magnetic 
pattern, changing the structural distortion has very little effect on the DOS away from the Fermi energy.  
There were no large shifts of spectral weight to lower energies, though small shifts of the order of 0.05 
eV did occur and these were within 0.5 eV of the Fermi energy (for comparison see the heavy restructuring 
of the DOS in Figs. \ref{DOS122} and \ref{FETEDOS}). The distortion can therefore be treated as a linear 
perturbation with a one-electron energy lowering observable at (or very near) the Fermi energy.  
Specifically, we find that the lowest energy structure corresponds to the smallest Fermi surface area.  
As an example, in Fig. \ref{FS122} we show the Fermi surface in the magnetic Brillouin zone of the fully 
relaxed (lowest energy) structure and a 'reverse distortion' in which the distances between like and 
unlike spins are reversed from the correct configuration.  The change in the size of the Fermi surface is 
clearly visible.  We were unable to engineer a further minimization of the Fermi surface with any choice 
of in-plane distortions other than the optimal energy one.

In conclusion, we have shown that the relevant physics with respect to the magnetic ordering and structural 
distortion in the ferropnictides lies in the one-electron energies.  Our results resolve the superficially 
binary choice between superexchange and Fermi surface nesting in favor of a third mechanism that is neither 
fully localized nor fully itinerant.  One-electron energy is gained throughout an energy range of at least 1 
eV below E$_F$ and the ground state is determined by which magnetic pattern most effectively exploits a 
downshift in spectral weight, not by fermiology. On the other hand, the Fermi surface itself is the 
operative feature for determination of the structural distortion.  The energy minimum for an in-plane 
distortion corresponds to a simultaneous minimization of the Fermi surface area.

We thank H. Eschrig, K. Koepernik and T. Yildirim for useful and engaging discussions related to this work.  We acknowledge funding from the Office of Naval Research.


\begin{thebibliography}{9}
\bibitem {note1} We are interested in how the double stripe structure manifests in stoichiometric FeTe 
within
LDA.  We cannot make meaningful comparisons with experiment and we therefore do not fully relax the FeTe 
structure, but only
relax the internal positions.  As before, relaxations are done separately for magnetic and nonmagnetic
cases.  We also neglect the checkerboard case as being of only secondary interest for this system.
\end{thebibliography}
\end{document}